\documentclass[preprint,12pt]{elsarticle}

\usepackage{graphicx}
\usepackage{epstopdf}
\usepackage{amsmath}
\usepackage{amssymb}
\usepackage[version=4]{mhchem}
\usepackage{siunitx}
\usepackage{longtable,tabularx}
\usepackage{fancyvrb}
\usepackage{color}
\usepackage{tcolorbox}
\definecolor{navyblue}{rgb}{0.0, 0.0, 0.5}
\usepackage[framemethod=TikZ]{mdframed}
\mdfdefinestyle{MyFrame}{%
	linecolor=gray,
	outerlinewidth=0.3pt,
	roundcorner=3pt,
	innertopmargin=5pt,
	innerbottommargin=5pt,
	innerrightmargin=5pt,
	innerleftmargin=5pt,
	frametitlerule=true,
	frametitlebackgroundcolor=white!80!gray,
	backgroundcolor=white!100!gray}
\definecolor{royalblue(traditional)}{rgb}{0.0, 0.14, 0.4}
\definecolor{sapgreen}{rgb}{0.31, 0.49, 0.16}
\definecolor{blue-green}{rgb}{0.0, 0.87, 0.87}
\definecolor{capri}{rgb}{0.0, 0.75, 1.0}
\definecolor{bleudefrance}{rgb}{0.19, 0.55, 0.91}
\usepackage{float}

\journal{}

\begin{document}

\begin{frontmatter}



\title{Verification Framework for Control System Functionality of Unmanned Aerial Vehicles}


\author{Omar A. Jasim   \      and  \	Sandor M. Veres}

\address{Department of Automatic Control and Systems Engineering, University of Sheffield, Amy Johnson Building, Mappin Street, S1 3JD, UK \\ Email:\{oajasim1@sheffield.ac.uk  ,  s.veres@sheffield.ac.uk\}}

\begin{abstract}
A control system verification framework is presented for unmanned aerial vehicles using theorem proving. The framework's aim is to set out a procedure for proving that the mathematically designed control system of the aircraft satisfies robustness requirements to ensure safe performance under varying environmental conditions. Extensive  mathematical derivations, which have formerly been carried out manually, are checked for their correctness on a computer. To illustrate the proceedures, a higher-order logic interactive theorem-prover and an automated theorem-prover are utilized to formally verify a nonlinear attitude control system of a generic multi-rotor UAV over a stability domain within the dynamical state space of the drone. 
Further benefits of the proceedures are that some of the resulting methods can
be implemented onboard the aircraft to detect when its controller breaches its flight envelop limits due to
severe weather conditions or actuator/sensor malfunction. Such a detection procedure can be used to advise the remote pilot or an onboard intelligent agent to decide on some alterations of the planned flight path or to perform emergency landing.

\end{abstract}




\end{frontmatter}


\section{Introduction}
\label{intro}
Control law design for aircraft  normally combines control engineering knowledge with tests of stability and smoothness of control responses under disturbances. This often takes the form of an iterative process of remodelling aerodynamics in wind tunnels and ultimately in flight tests. Given a particular open loop dynamical model, control engineering relies on mathematical theory that is implemented in computations of flight controllers onboard. When the flight envelop is defined, it introduces numerical values which need to be carried through derivations and proofs of stability and acceptable handling within the flight envelop. In this paper manual derivations are replaced by theorem proving methods. The procedures presented go  beyond algebraic  computation and use higher-order logic, including handling of functionals, operators, concepts of convergence, stability and levels of smoothness measures. 

The approach taken in this paper fits into one of  the three stages of formally verifiable controller design as outlined in \cite{c47}, where robust control theory verification is  followed by verification of the software used for implementation. A third stage is to prove that the quantisation effects of the obtained digital controller do not significantly effect the results in stage one of the controller theory used.   In implementations of aviation software, verification is often followed by redundancy-based safety analysis for critical sensors and actuators using voting principles, the affect of which lies  outside the scope of this paper. 

Realtime code-verification is systematically checking the correctness of the encoded controller  such as in \cite{c50,c48} and \cite{c49}. This paper focuses on verification of the control principles before it is implemented in realtime code. An interactive theorem prover (ITP) is used for performance verification under nominal environment conditions and onboard stability and performance monitoring is based on an automated theorem prover (ATP) for excessive conditions, including some sensor or actuator failures. 

There have been prior initiatives  on   verification of safety-critical and cyber-physical systems. Such are the European Integrated Tool Chain for Model-based Design of Cyber-Physical Systems (INTO-CPS)\cite{c56}, where a Functional Mockup Interface
(FMI) has been developed for integrating the formal verification of Cyber-Physical Systems using the PVS \cite{c7} theorem prover  with model-based software to co-simulation these systems. This approach  integrates simulated models in model-based tools such as Modelica\cite{FV_mod} , Simulink/Matlab \cite{FV_matlab} or 20-sim \cite{FV_20_sim} with the FMI interface to verify the control system according to the required specifications using formal methods. 

The FMI is also implemented in \cite{c88i} using Isabelle/UTP \cite{c73}, where Modelica is used to model the control system of a train then the model is encoded in Isabelle/UTP with FMI framework for co-simulation. Another project is the ERATO Metamathematics for Systems Design (MMSD) \cite{c57}, where a framework is developed to use formal methods to verify automotive-related applications in industry such as cars. Other projects are conducted by the verification team of NASA Langley Research Center \cite{c70} such as \cite{c96,c97} and \cite{c98}. Though   these approaches  use formal methods to verify control systems, the derivations of control laws are not covered by verification before implementing them. In addition, there is an absence of onboard real-time stability monitoring of these systems by formal methods.    

There have been several studies conducted to formally verify control systems. In \cite{c72}, a framework for reasoning about the transfer function and steady state errors in feedback control systems using HOL Light \cite{c74} theorem prover   is proposed. Adnan, et al. \cite{c65} formalised the theory of  Laplace transforms using interactive theorem prover to analyse linear control systems. In \cite{c69}, formal verification of transformation matrices of a coordinate system is proposed for aircraft control systems using the Coq \cite{c6} theorem prover.  An autonomous vehicle controller is formalised and proved in \cite{c63} using the PVS \cite{c7} theorem prover to illustrate its correctness in terms of stability of the Jacobian matrix and then co-simulated using the PVSio environment in PVS prover. Denman et at. \cite{c27} presented a method of implementing a formal Nichols plot analysis of a flight control
system using the MetiTarski ATP \cite{c5} for stability verification. Araiza-Illan et al. \cite{c29} proposed a framework for automatic translating the system block diagrams modelled in Simulink into the Why3 \cite{c21} prover for verification. Verifying hybrid control systems using differential dynamic logic in KeYmaera \cite{c77} prover also proposed in \cite{c75,c76} where abstraction of the closed-loop behaviour is constructed for a class of controllers.   

Our proposed framework is different from this approach as we are verifying the correctness of the derived control law of aviation systems at the design stage before the simulation step  using an  ITP, followed by   real-time monitoring of  stability by ATP. 

The novelty of this paper is to  demonstrate that formal methods in some iterative  theorem provers  are suitable to verify and prove the correctness of robust control theory for a prescribed flight envelope of a multirotor aerial vehicle. Although prior work suggested this may be a possibility, this is a first evidence of this kind. This involves  formal stability analysis to guarantee system's robustness then ensure aircraft's safety by conducting continuous onboard stability monitoring using interactive and automated theorem provers.  
The methods are implemented  in Isabelle \cite{c1} and MetiTarski \cite{c5} , and the codes have been made available online. 

The proposed verification framework is applied to a control scheme of a generic quad-copter presented in \cite{c88f}, which consists of a nonlinear attitude controller to deal with  modelling uncertainty and external disturbances that is designed using the well-known dynamic inversion technique \cite{c54,c43}.  Lyapunov's method \cite{c39} has been used as part of the attitude controller design and to analyse the system to ensure  its stability. We have applied our verification scheme to the attitude control system since all definitions, assumptions, derivations and performance proofs of the system are formalised then proved in the Isabelle higher-order logic (HOL) ITP. Lyapunov stability of the control system is formalised and proved in Isabelle/HOL. We have implemented the stability analysis into the MetiTarski ATP for onboard stability testing to check whether the aircraft is in its stable region or goes beyond it since in the latter case the autopilot system can inform the pilot/station to perform an urgent action. 

The paper is structured as follow: section 2 describes theorem proving tools where Isabelle and MetiTarski provers are introduced; section 3 presents the proposed verification framework of unmanned aerial vehicles; section 4 demonstrates the applicability of the framework on a multirotor control system and describe the onboard stability monitoring; and section 5 presents the conclusions of this work.

\section{Theorem Proving}

Computer based theorem proving is a computational tool set in some logical system that can be used to prove the soundness and correctness of  mathematical arguments. There are two main types of theorem proving: Automated
Theorem Proving (ATP), which proves mathematical statements automatically  without human interaction, and Interactive Theorem Proving (ITP), which automates steps of 
formal proofs by aid of a developer guiding the process of proof. The automated steps rely on mathematical logic and automated reasoning techniques. 
ITP  will be used and described further in this paper for  the verification framework proposed. ITPs are proof assistants, which  formally define and prove mathematical theorems. Therefore, their user can
implement a mathematical theory by defining assumptions and some valid
logical statements, to start with. Then the ITP procedure will try to prove a sequence of statements,  relying on available formal theories, and also by using existing logical methods and techniques or some external resources, such as ATP tools . 

The distinction between ITP and ATP systems is not only that ATP systems fully automate proofs but that ATPs tend to have restricted
expressivity where, unlike ITPs, they cannot prove higher order mathematical theories. Instead, they can be used to prove non complex mathematical formulas that contain inequalities over real numbers, quantified variables, and special functions. On the other hand, ITP systems have the ability to support formalising and proving  mathematical theories, which involve higher order logic,  with the aid of a human supervisor in an interactive way.

However, using ATPs can be
utilised locally in an ITP to prove a step in the proof,  by adding their packages to the  ITP.   This can also be done online 
by using System on TPTP \cite{c88d}, which is a web-based system including
the most powerful ATPs that can be used to proof mathematical statements automatically.
A good example of a useful ATP, from the control engineering side, is
\textit{MetiTariski} , which is a first-order logic (FOL) prover that was designed to work over the field of real numbers.

One of the well known tools in testing of the designed models and formally
verifying and validating their correctness are model checkers \cite{c18}. These tools are used to
model a system as a finite state transition, like automata or timed
automata, and system properties are expressed in the form of proposition
temporal logic. Then, the checking problem is reduced to a graph search and an
exhaustive exploration of all possible states is accomplished using symbolic
algorithms. However, model checkers are not applicable with infinite state space designs and they can not use to prove the derivation correctness of the designed control system. In contrast, ITPs can be employed to verify infinite state space systems and mathematical arguments.    

\subsection{Isabelle/HOL}
Isabelle is an interactive theorem prover (proof assistant) based on automated reasoning techniques and logical calculus. It is written using the  ML   functional programming language such that the rules are presented as propositions not as functions and the proofs are structured by combining rules based on the $\lambda$-calculus. 
It provides the ability to express the mathematical formulae in a formal language and prove them using different logical tools. 
Isabelle/HOL is one of Isabelle's platforms, which is based on higher-order logics with quantifiers and semantics. Isabelle/HOL has a proof language called \textit{Isar} which structures the proofs in a readable and understandable mathematics-like syntax for both users and computers. The mathematical formulas can be formalised and proven in the Isar language with the aid of Isabelle's logical tools. Examples of such tools are the
\textit{simplifier}, which performs operation and reasoning on equations, the \textit{classical reasoner} that carry out long chains of reasoning procedures to prove statements or theories, automatic proof of \textit{linear arithmetic} statements, \textit{algebraic decision procedures} for decision making verification, advanced \textit{pattern matching}, and \textit{sledgehammer} for automatically finding the proofs based on already proven theorems in Isabelle's library and also calling external FOL provers (ATPs).

However, Isabelle has been chosen in this work for its rich logical and automation techniques since it has a
large library, which contains most of the mathematical theories that are useful in control systems verification. 
Some examples of these theories are real, integer and complex numbers; Euclidean, vector, norm spaces; differentiation and integration; real and transcendental norms; higher-order
functions on different of number types; logical techniques; algebra; etc. Isabelle also support other logical systems such as Hoare logics for systems verification \cite{c14} in addition to the ability to generate executable codes of the proved statements.

There is a wide range of syntax and command types in the Isabelle/HOL, therefore, the most
common and useful one for our framework  will be described. Isabelle/HOL expressions and
symbols are described in Table \ref{Tab1}.

\begin{table}
	\caption{Isabelle/HOL symbols and expressions}
	\label{Tab1}       
	\begin{center}
		\begin{tabular}{ll}
			\hline\noalign{\smallskip}
			Expression & Description \\ 
			\noalign{\smallskip}\hline\noalign{\smallskip}
			$\rightarrow$  & \multicolumn{1}{m{10cm}}{mapping from value to value or function to function.}   \\
			$\longrightarrow$  & \multicolumn{1}{m{10cm}}{refers to "\textit{imply}" in HOL.}   \\
			$\Rightarrow$  & \multicolumn{1}{p{10cm}}{used to define a function with its corresponding  variables types (e.g.,
				$real$ $\Rightarrow$ $real$) that is a function maps from real to real variable.}  \\			
			$\implies$  &   \multicolumn{1}{p{10cm}}{refers to "\textit{imply}" in \textit{Isar} language in Isabelle. (e.g., $x=0$ $\implies$ $y=x$) that
				is x=0 is an assumption and y=x is the statement to be proven.} \\	
			$\bigwedge$  &   \multicolumn{1}{p{10cm}}{refers to "\textit{for universal all}" which applies to all assumptions and/or proof statements.} \\	
			$\forall $  &   \multicolumn{1}{p{10cm}}{means "\textit{for all}" or "\textit{for any}" and it is for a specific assumption or statement.} \\	
			$\exists$  &   \multicolumn{1}{p{10cm}}{means "\textit{there exist}" or "\textit{there is}".} \\
			$\exists!$  &   \multicolumn{1}{p{10cm}}{means "\textit{there is only one}".} \\		
			$\wedge$  &   \multicolumn{1}{p{10cm}}{refers to the logical “\textit{and}”.} \\	
			$\lor $  &   \multicolumn{1}{p{10cm}}{refers to the logical “\textit{or}”.} \\	
			$x^{'}$  &   \multicolumn{1}{p{10cm}}{the $1^{st}$ time derivative of $x$ ($x^{''}$ the $2^{nd}$ time derivative of $x$).} \\		
			$|x|$  &   \multicolumn{1}{p{10cm}}{refers to absolute value of $x$.} \\
			$x\$i$  &   \multicolumn{1}{p{10cm}}{returns the $i^{th}$ element of the vector $x$.} \\
			$\bullet$  &   \multicolumn{1}{p{10cm}}{an operator for the \textit{dot product} of two vectors.} \\
			$*_v$  &   \multicolumn{1}{p{10cm}}{an operator for the multiplication of a matrix and a vector.} \\
			$*_s$  &   \multicolumn{1}{p{10cm}}{an operator for the multiplication of a scalar value and a vector.} \\
			$**$  &   \multicolumn{1}{p{10cm}}{an operator for the multiplication of two matrices.} \\
			$(\lambda \  t. \ x \ t)$  &   \multicolumn{1}{p{10cm}}{this is equivalent to the  function $x(t)$ but under the constraint of an
				argument ($t$).} \\	
			$norm(x)$  &   \multicolumn{1}{p{10cm}}{the Euclidean norm of a vector or a matrix.} \\
			$SUP(x)$  &   \multicolumn{1}{p{10cm}}{the supremum value of $x$.} \\
			\noalign{\smallskip}\hline
		\end{tabular}
	\end{center}
\end{table}

\subsection{MetiTarski}
MetiTarski is a FOL automated theorem prover, which is designed as a combination of the resolution prover $Metis$ \cite{c55} and a decision procedure tool $QEPCAD$ \cite{c52} to conduct proofs over the field of real numbers. It can be utilised to prove universally quantified inequalities of transcendental and other functions,  for instance $ln$, $log$, $exp$, etc. MetiTarski reduces the problems to be decidable by replacing the functions with upper and lower bounds on the real numbers field. It can perform proofs with the aid of three reasoners: QEPCAD, Z3 \cite{c53} and Mathematica. However, we have chosen MetiTarski to check the stability of  unmanned aerial vehicle (UAV) systems due to its features of proving quantified inequalities including the above functions over large scale real numbers.

\section{An Aircraft Verification Framework}
\begin{figure}[thpb]
	\centering
	\includegraphics[scale=0.5]{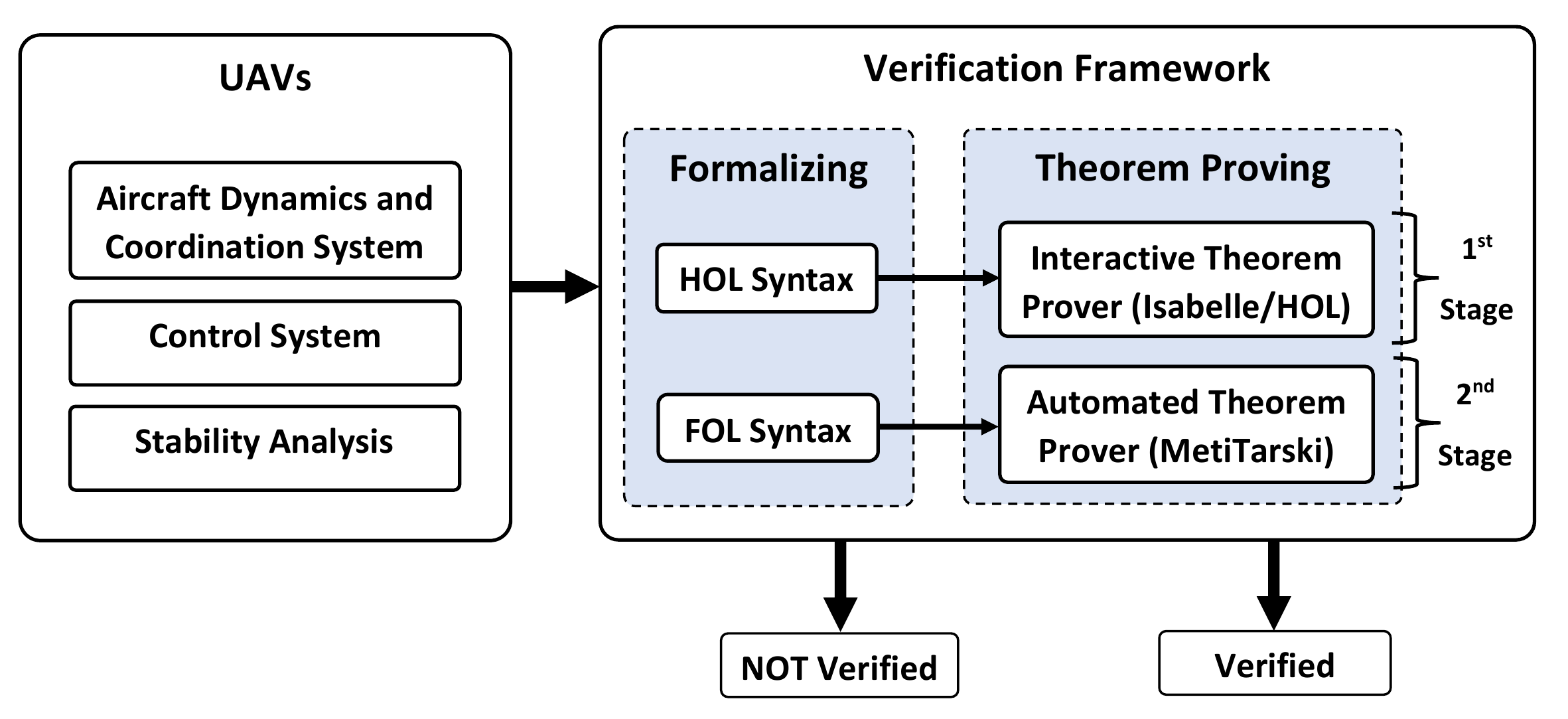}
	\caption{UAVs verification framework}
	\label{fig1}
\end{figure}

The proposed verification framework is shown in Fig.\ref{fig1} which starts with formalising the aircraft's dynamical equations of motion, the coordinate system in the rigid body frame, the controller design, and stability analysis in the HOL syntax of the Isabelle prover. In addition, the stability analysis step  is formalised in the FOL syntax of the MetiTarski ATP for possible verification of stability onboard the craft.   

The verification framework consists of two stages: the ITP represented by \textit{Isabelle/HOL} to prove the mathematical derivation of the designed control system and its stability analysis, and the ATP represented by \textit{MetiTarski} prover for continuously checking the validity of aircraft's stability onboard during the flight. The next section will illustrate these two verification stages in detail.       

\section{Case Study: Multirotor Verification}
\subsection{Verification in Isabelle/HOL Prover}
In this subsection, we will demonstrate the first stage of the verification framework shown in Fig. \ref{fig1} using an attitude controller of a generic quadcopter UAV proposed in \cite{c88f} with considering the assumptions and flight conditions made. 
In this proposed controller, the quadcopter's rotational dynamics are controlled using a robust nonlinear controller that takes into account the modelling uncertainty and external disturbances.  To ensure correctness of the designed attitude control where simulation cannot guarantee that the UAV's control system is robust for all possible flight conditions, the design's derivations and stability analysis have been verified using Isabelle/HOL prover.  
Isabelle/HOL is chosen for this purpose due to its rich library of mathematical theorems which are required to perform the UAV's control system verification. 

\begin{figure}[thpb]
	\centering
	\includegraphics[scale=0.3]{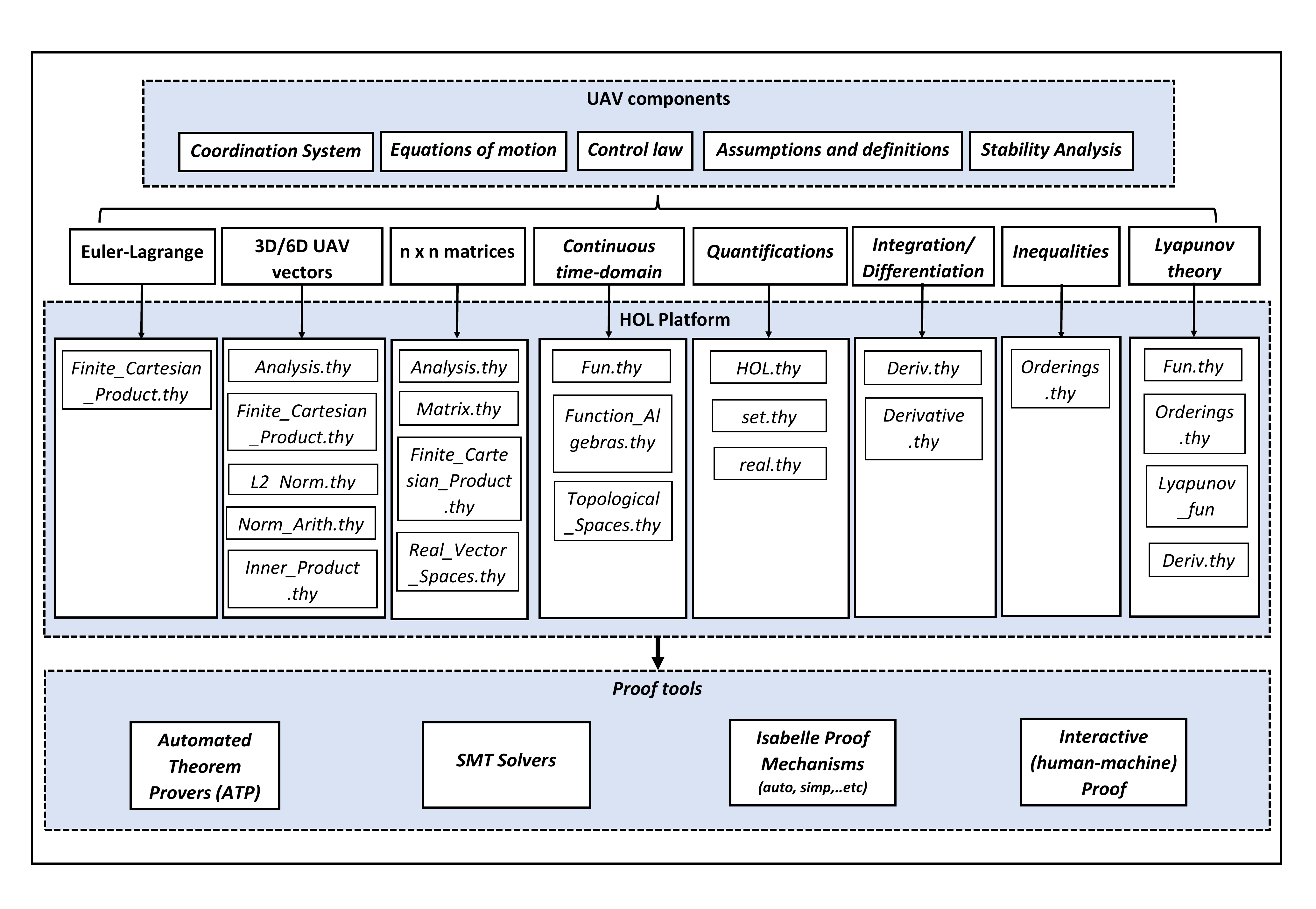}
	\caption{Formalising and proving UAV's controller in Isabelle/HOL theorem prover}
	\label{fig3}
\end{figure}

The verification process using Isabelle/HOL is illustrated in Fig. \ref{fig3} which mainly consist of two stages: formalising and proving procedures. The first stage starts by formalising the quadcopter UAV system into Isabelle/HOL syntax such as  the coordinate system, rotational dynamics, time-domain functions, proposed assumptions and aircraft's stability analysis. The implementation of the control design and aircraft's dynamics includes a series of \textit{definition}, \textit{lemma}, and \textit{theorem} items. Some assistant lemmas were needed to be formalised and proven, which did not exist in Isabelle due to that prover library is still under development as is the case  with other theorem provers. The formalisation also needs to import some pre-proven mathematical theories and lemmas from the prover library, which are used in formalising and proving the control system equations with the proposed assumptions and definitions. 

Much needed theories under $HOL$ platform will be described here to illustrate the formalisation and proof procedures. First of all, 
the main multi-variable analysis package which includes $Analysis.thy$ for functions operations over real field, 
\textit{Finite\_Cartesian\_Product.thy} and \textit{Inner\_Product.thy} for definitions and operations of real vectors,  \textit{L2\_norm.thy} and \textit{Norm\_Arth.thy} for real vector norms and their operations, etc. The $HOL.thy$ is the core of $HOL$ platform which includes definitions of real numbers (\textit{real.thy}), functions (\textit{fun.thy}), sets (\textit{set.thy}), etc., which are necessary in all the formalising procedures. 
The main multi-variable analysis theory, $Analysis.thy$, which includes definitions of real vectors (\textit{Finite\_Cartesian\_Product\\ .thy}), vector norms   (\textit{L2\_norm.thy, Norm\_Arth.thy}) and their operations. These theories are used to define the aircraft's three-dimensional rotation vectors and their norms such as the torque, angular velocity and acceleration vectors where each component of a vector represented as a continuous time-domain function $f(t)$; the continuous function defined in $Fun.thy$, $Function\_Algebras.thy$ and $Topological\_Spaces$  theories are utilised for this purpose. The time sub-domain is defined by a time set  $T= \{t. \ t \in \{0..\infty\}\}$ and is followed by definitions of sets of vectors, which are working within  $T$.

The matrices components are formalised using $Matrix.thy$ and their operations using $Analysis.thy$, $Finite\_Cartesian\_Product.thy$ and $Real\_Vector$ $\_Spaces.thy$. The rate change of the quadcopter attitudes, i.e. velocities and accelerations, are formalised by time derivation using $Deriv.thy$ and $derivative.thy$ theories. The quadcopter controller design includes several robust assumptions which need inequalities over the real-numbers field. Fortunately, such inequalities have been defined in Isabelle prover under $Orderings$ $.thy$ theory. This is an important feature for any robust control design to be proven. However, the second stage (proving procedures) is an interactive process between the designer/engineer and the automated proving tools that Isabelle prover has or supported. The role of designer/engineer is to help the prover to step-by-step prove the statement in case that the prover is not able to solve the proof automatically by simplifying the statement into several steps. Each step should be proven suing the provided automated tools before moving to the next one otherwise the prover will not pass the statement. Example of the automated tools supported in Isabelle are: \textit{CVC4}, \textit{Z3}, \textit{SPASS}, \textit{E  prover}, \textit{Remote\_Vampire} and \textit{SMT sovlers}. In addition, Isabelle has its own automatic proving tools such as \textit{auto}, \textit{simp}, \textit{blast}, etc.  
Most of the control system that we have verified required our interaction with the prover due to the design complexity for the prover to solve them automatically.      
The Isabelle code is long to be stated here, where only the important definitions  and proofs are shown, while the complete code can be found in our online  repository\footnote{\textit{https://github.com/Formal-Methods-of-Robotics/Quadcopter-veri}}.

The quadcopter attitude dynamics in Eq. 5 of \cite{c88f} is,
\begin{equation} \label{FV_eq40b}
J(\eta) \ddot{\eta}  +  C(\eta,\dot{\eta}) \dot{\eta} + d = \tau , 
\end{equation}
which formalised in Isabelle/HOL as can be seen the following code 
\begin{mdframed}[style=MyFrame, frametitle 	={}]	
	\begin{scriptsize}	
		\textcolor{royalblue(traditional)} {\textbf{definition}} $"$\textcolor{blue} {\textbf{$att\_dyms$}} $\tau \ J \ C \ \eta \ \phi \ \theta \ \psi  \ d = 
		(\forall t\in T. \ (\forall \eta' \ \eta''. \ \eta \in D3\_vec\_set) \wedge Euler\_vec \ \eta \ \phi \ \theta \ \psi \wedge
		(\forall i.((\lambda t. \ \eta\$i) \ has\_derivative (\lambda t. \ \eta'\$i)) (at \ t \ within \ T)) \wedge J\_mat \ J \ \phi \ \theta \ \psi \wedge \\
		(\forall i.((\lambda t. \ \eta'\$i) \ has\_derivative (\lambda t. \ \eta''\$i)) (at \ t \ within \ T)) \wedge C\_mat \ C \ \phi \ \theta \ \psi \wedge  \\ \tau = J *_v \eta'' + C *_v \eta' + d)"$ 
	\end{scriptsize}
\end{mdframed} 
and the torque vector $\tau$ in Eq. 9 - \cite{c88f}, 
\begin{equation}  \label{FV_eq45}
\tau =
\begin{bmatrix}
\tau_{\phi} \\ \tau_{\theta} \\ \tau_{\psi}
\end{bmatrix}
=
\begin{bmatrix}
\ell k(\omega^{2}_{2} - \omega^{2}_{4}) \\ 
\ell k(- \omega^{2}_{1} + \omega^{2}_{3}) \\ 
b(-\omega^{2}_{1} + \omega^{2}_{2} - \omega^{2}_{3} + \omega^{2}_{4})
\end{bmatrix},
\end{equation}
is defined by bounding all propeller angular velocities $\omega_i$ with their maximum value $\omega_{max}$ as
\begin{mdframed}[style=MyFrame, frametitle 	={}]	
	\begin{scriptsize}	
		\textcolor{royalblue(traditional)} {\textbf{definition}} $"$\textcolor{blue} {\textbf{$torq\_fun$}} $\tau = ((\exists \ \omega_1 \ \omega_2 \ \omega_3 \ \omega_4. \mid\omega_1\mid<\omega_{max} \wedge \mid\omega_2\mid<\omega_{max}  \wedge
		\mid\omega_3\mid<\omega_{max} \wedge \\  \mid\omega_4\mid<\omega_{max} \wedge
		\tau \in D3\_vec\_set \ \wedge  \\
		\tau\$1= \ell*k*(\omega^{2}_2 - \omega^{2}_4) \ \wedge \
		\tau\$2= \ell*k*(-\omega^{2}_1 + \omega^{2}_3) \  \wedge \
		\tau\$3= b*(-\omega^{2}_1 + \omega^{2}_2 - \omega^{2}_3 + \omega^{2}_4) ))"$ 
	\end{scriptsize}
\end{mdframed}   

\noindent The control input $u$ in Eq. 18 - \cite{c88f},
\begin{equation} \label{FV_eq17}
\begin{split}
u = \ddot{\eta}_{d} + K_{r}\dot{e} + K_{\eta}e 
\end{split},
\end{equation} 
and the control law in Eq. 12 - \cite{c88f},
\begin{equation} \label{FV_eq8}
\tau = \hat{J}(\eta) u  +  \hat{N}(\eta,\dot{\eta}) + \hat{d} +\gamma,
\end{equation}
are  defined in the prover as the following code, 
\begin{mdframed}[style=MyFrame, frametitle 	={}]	
	\begin{scriptsize}	
		\textcolor{royalblue(traditional)} {\textbf{definition}} $"$\textcolor{blue} 
		{\textbf{$cont\_u$}} $(u :: (real, 3) vec) \ \eta_d'' \ K_r \ K_{\eta} \ e  \ e' \ \  = \ \
		(u= \eta_d'' + K_r *_v e' + K_{\eta} *_v e)"$ \\
		
		\noindent \textcolor{royalblue(traditional)} 
		{\textbf{definition}} $"$\textcolor{blue}
		{\textbf{$cont\_law$}} $(\tau :: (real, 3) vec) \ u \ \gamma \ \ J_{hat} \ N_{hat} \ d_{hat}\  = \ \ (\tau = J_{hat} \ *_v \ u + N_{hat} + d_{hat} + \gamma)"$ 		 
	\end{scriptsize}
\end{mdframed} 

\noindent The derivation in Eq. 19 - \cite{c88f},
\begin{equation} \label{FV_eq18}
\begin{split}	
\ddot{\eta} &= \hat{J}(\eta) J^{-1}(\eta) u  + J^{-1}(\eta) [\Delta N(\eta,\dot{\eta}) + \Delta d] \\
& \ \ \ + J^{-1}(\eta) \gamma  \\
&= u + (\hat{J}(\eta) J^{-1}(\eta) - I) u + J^{-1}(\eta) [\Delta N(\eta,\dot{\eta}) + \Delta d]  \\
& \ \ \ + J^{-1}(\eta) \gamma \\
& = u - v + J^{-1}(\eta) \gamma \\
where \\
v &=[I - \hat{J}(\eta) J^{-1}(\eta)] u  -J^{-1}(\eta) [\Delta N(\eta,\dot{\eta}) + \Delta d].
\end{split}
\end{equation}
is formalised and proved in Isabelle/HOL based on the $att\_dyms$, $cont\_u$ and $cont\_law$ (see the proof in the repository). The closed-loop error dynamic in Eqs. 22 and 23 of \cite{c88f},
\begin{equation} \label{FV_eq20}
\begin{split}
\dot{E} &=  AE + B[v - J^{-1}(\eta) \gamma]
\end{split} 
\end{equation}
where
\begin{equation} \label{FV_eq20a}
A = 
\begin{bmatrix}
0^{3 \times 3} & I^{3 \times 3} \\
-K_{\eta}^{3 \times 3} & -K_{r}^{3 \times 3} 
\end{bmatrix} 
, \ B = 
\begin{bmatrix}
0^{3 \times 3} \\ I^{3 \times 3}
\end{bmatrix},
\end{equation}
are implemented as 
\begin{mdframed}[style=MyFrame, frametitle 	={}]	
	\begin{scriptsize}	
		\textcolor{royalblue(traditional)} {\textbf{lemma}} $ Eq\_22: \\
		$\textcolor{sapgreen} {\textbf{assumes}}  $"\forall t. \ t\in T"$\textcolor{sapgreen} {\textbf{and}}$"(set\_of\_definitions \ \ \phi \ \theta \ \psi \ \phi_d \ \theta_d \ \psi_d \ \eta \ \eta_d \ \eta' \ \eta_d' \ \eta'' \  \eta_d'' \ u \ \gamma \ e \ e' \ e'' \ \tau \ d \ d_{hat} \\ E \ v \ \delta \  C \ C_{hat} \ 
		N \ N_{hat} \ \Delta N \ \Delta d \ A \ B \ W \ Q \ P \ K_{\eta} \ K_r \ J \ J_{hat}"$ \\   
		\textcolor{sapgreen} {\textbf{shows}} $ "E' = A *_v E + B *_v (v - (matrix\_inv(J) *_v \gamma))" \\
		$\textcolor{royalblue(traditional)} {\textbf{proof -}}$ \\
		$\textcolor{royalblue(traditional)} {\textbf{have}} \ \ $ "e'' = \eta_d'' - \eta''" 
		$ \ \ \textcolor{royalblue(traditional)} {\textbf{using}} \ \ $ assms \ \ ddot\_error\_vec\_def \ \ set\_of\_definitions\_def  \ \ $\textcolor{royalblue(traditional)} {\textbf{by}} $ \ metis \\
		$\textcolor{bleudefrance} {\textbf{thus}} \ \ $ ? \ thesis \ 
		$\textcolor{royalblue(traditional)} {\textbf{by}} $ (smt \ \ B\_mat\_def \ \ assms(2) \ \ exhaust\_3 \ \ set\_of\_definitions\_def)  \ \
		$\textcolor{royalblue(traditional)} {\textbf{qed}}$ $		 
	\end{scriptsize}
\end{mdframed}    
\

\noindent 
The $set\_of\_definitions$ in the above code is a definition used to call all the pre-defined definitions in one step. We formalised the assumptions proposed in \cite{c88f} as follow: 

\noindent 
$Assumption \ 1$ (Eq. 14 \cite{c88f}),
\begin{equation} \label{FV_eq21a}
\| \Delta d \| \le D, \ \ \| d \| + D <\bar{D}
\end{equation}
$Assumption \ 2$ (Eq. 15 \cite{c88f}),
\begin{equation} \label{FV_eq12b}
\| \Delta N(\eta,\dot{\eta}) \| \le S .
\end{equation} 
$Assumption \ 3$ (Eqs. 24, 25 and 26 \cite{c88f}),
\begin{equation} \label{FV_eq21}
sup (\| \ddot{\eta}_{d}\|) < H  
\end{equation}
\begin{equation} \label{FV_eq22}
\| I - \hat{J}(\eta) J^{-1}(\eta) \|  \le \xi \le 1,
\end{equation}
\begin{equation} \label{FV_eq22a}
\beta_{min} \le \| J^{-1}(\eta) \| \le \beta_{max}, 
\end{equation}
in Isabelle/HOL as follow:
\begin{mdframed}[style=MyFrame, frametitle 	={}]	
	\begin{scriptsize}	
		\textcolor{royalblue(traditional)} {\textbf{definition}} $"$\textcolor{blue} {\textbf{$assump1$}} $(d::(real,3)vec) \ d_{hat} \  \Delta d  \ = \ (\Delta d = d_{hat}-d \ \wedge \   norm(d)\le D \ \wedge \ norm(d)+D<D_{bar})"$ \\
		\textcolor{royalblue(traditional)} {\textbf{definition}} $"$\textcolor{blue} {\textbf{$assump2$}} $(N ::(real,3)vec) \ N_{hat} \ \Delta  N \  = \ (\Delta N = N_{hat}-N \ \wedge \   norm(\Delta N)\le S)"$	\\	
		\textcolor{royalblue(traditional)} {\textbf{definition}} $"$\textcolor{blue} {\textbf{$assump3\_a$}} $\eta_{d}'' \  = (($SUP$\ t\in T. \ \ norm(\eta_{d}'')) < H)"$	\\
		\textcolor{royalblue(traditional)} {\textbf{definition}} $"$\textcolor{blue} {\textbf{$assump3\_b$}} $ J \ J_{hat} \ \phi \ \theta \ \psi \ = \ (J\_mat \ J \ \phi \ \theta \ \psi \ \wedge \ J_{hat}\_mat \ J_{hat} \ \phi \ \theta \ \psi \ \wedge \ norm(mat \ 1 - J_{hat} \ ** \ matrix\_inv(J)) \le \xi \wedge \  \beta_{min}  \le  norm(matrix\_inv(J)) \ \wedge \ norm(matrix\_inv(J)) \ \le \ \beta_{max})"$
	\end{scriptsize}
\end{mdframed}
\

Stability analysis of the attitude controller as stated in Eqs. 27-37 of \cite{c88f} is implemented in Isabelle/HOL using a set of definitions, $definition$, several lemmas, $lemma$, and short theorems in terms of theorem,  $theorem$. This structure of using several \textit{lemmas} and \textit{theorems} during the proof is due to the fact that the reasoning system of the theorem prover cannot handle long proofs with many assumptions, i.e. the system unable to  prove many equations if they are formalised in only one \textit{lemma} or \textit{theorem} style. However, the stability analysis starts by defining the candidate Lyapunov function $V$ as in Eq. 27 of \cite{c88f} 
which is formalised as a \textit{definition} in Isabelle/HOL:
\begin{mdframed}[style=MyFrame, frametitle 	={}]	
	\begin{scriptsize}	
		\textcolor{royalblue(traditional)} {\textbf{definition}} $"$\textcolor{blue} {\textbf{$Lyapunov$}} $V \ E = (\forall t\in T. \ \  \textbf{\textit{if}} \ (E :: (real,6) vec) \ne 0  
		\ \ \textbf{\textit{then}} \ \ (\exists a.\  V(E)= (a:: real) \ \ \wedge \ \ continuous\_on \ \  D6\_vec\_set \ V  \ \ \wedge \ V(E)>0) \ \ \textbf{\textit{else}} \ \ V(E) = 0)"$		 
	\end{scriptsize}
\end{mdframed}

\noindent Taking the candidate Lyapunov function $V$, the time derivative of Lyapunov function is derived and the derivations in Eqs. 28-30 in \cite{c88f} are proven symbolically  and detailed in the online repository above. 
\begin{mdframed}[style=MyFrame, frametitle 	={}]	
	\begin{scriptsize}	
		\textcolor{royalblue(traditional)} {\textbf{theorem}} $Stb\_Eqs\_28\_30:$ \\
		\textcolor{sapgreen} {\textbf{assumes}}  
		$"\forall E. \ E \ne 0"$ 
		\textcolor{sapgreen} {\textbf{and}}
		$"Lyapunov \ V \ E"$ 
		\textcolor{sapgreen} {\textbf{and}}
		$"V(E) = E \bullet (Q *_v E)"$
		\textcolor{sapgreen} {\textbf{and}} 
		$"A\_mat \ A"$ \\
		\textcolor{sapgreen} {\textbf{and}}
		$"E' = A *_v E + B *_v (v - (matrix\_inv(J) *_v \gamma))"$	\\	
		\textcolor{sapgreen} {\textbf{and}} 
		$"(\forall t\in T. \ ((\lambda t. \ V(E)) \ \ has\_derivative \ \ (\lambda t. \ V'(E))) (at \ t \ within \ T))"$	\\  
		\textcolor{sapgreen} {\textbf{shows}}
		$"V'(E) = - (E \bullet (P *_v E)) + 2 * (((E \ v^{*} \ Q) \ v^{*} \ B) \bullet (v - matrix\_inv(J) \ *_v \ \gamma))"$ \\
		\textcolor{royalblue(traditional)} {\textbf{proof -}} 
		. . . 
		\textcolor{royalblue(traditional)} {\textbf{qed}}$ $		 
	\end{scriptsize}
\end{mdframed}        
\noindent  The term $\gamma$ in Eq. 31 of \cite{c88f}, 
\begin{equation} \label{FV_eq34}
\gamma =
\begin{cases}
\dfrac{ \delta(E)}{\|B^{T}Q E\|} B^{T}Q E  &\ \ \ \ \ \ \|B^{T}Q E\| \ge \sigma
\\
\dfrac{ \delta(E)}{\sigma} B^{T}Q E       &\ \ \ \ \ \ \|B^{T}Q E\| < \sigma
\end{cases}
\end{equation} 
is defined then the derivation in Eq. 32 - \cite{c88f} is performed using Cauchy-Schwartz inequality (see "$theorem$  $Eq\_32$" in the repository). Based on Eq. 33 - \cite{c88f},
\begin{equation} \label{FV_eq36}
\delta(E) \ge  \dfrac{\|v\|}{\beta_{min}}
\end{equation}
and the upper bound of norm of $v$ derived in Eq. 34 - \cite{c88f} (see "$theorem$ $Eq\_34$" in the repository), $\delta(E)$ in Eq. 35 - \cite{c88f} is obtained (see "$theorem$ $Eq\_35$" in the repository). The terms $\gamma$ and $\delta(E)$ are implemented in the prover as "$gamma$$\_ def$" and "$delta$$\_ def$" respectively,
\begin{mdframed}[style=MyFrame, frametitle 	={}]	
	\begin{scriptsize}	
		\textcolor{royalblue(traditional)} {\textbf{definition}} $"$\textcolor{blue} {\textbf{$gamma \_def$}} $\gamma \ B \ Q \ \delta \ E = (\forall t\in T. \ 
		\textbf{\textit{if}} (norm(transpose(B) *_v (Q *_v E)) \ge \sigma) 
		\ \textbf{\textit{then}} \\ (\gamma = (\delta/norm(transpose(B) *_v (Q *_v E))) *_s (transpose(B) *_v (Q *_v E))) 
		\ \textbf{\textit{else}} \\ (\gamma = (\delta/\sigma) *_s (transpose(B) \ *_v (Q *_v E))))"$	\\
		\textcolor{royalblue(traditional)} {\textbf{definition}} $"$\textcolor{blue} {\textbf{$delta\_def$}}
		$\delta \ (v::(real,3)vec) = (\forall \ t\in T. \ \exists \varepsilon . \ \varepsilon >0 \ \ \wedge \ \ norm(v) \le \varepsilon \rightarrow \ \delta \ge \varepsilon / \beta_{min})"$			 
	\end{scriptsize}
\end{mdframed} 
\noindent Note that the short arrow $\rightarrow$ in the code refers to \textit{implies} while the longer $ \longrightarrow$ refers to \textit{convergence} in HOL.

Finally, based on all the above definitions and assumptions, it has been verified that the proposed control system  is asymptotically stable since the time derivative of Lyapunov function in Eq. 36 and 37 - \cite{c88f},
\begin{equation} \label{FV_eq39}
\dot{V}(E) = - E^{T} P E + 2 E^{T}QB (v- J^{-1}(\eta) \dfrac{\delta(E)}{\|B^{T}Q E\|} B^{T}Q E)  <0 \ 
\end{equation} 
\begin{equation} \label{FV_eq39a}
\dot{V}(E) = - E^{T} P E + 2 E^{T}Q B(v- J^{-1}(\eta) \dfrac{ \delta(E)}{\sigma} B^{T}Q E )  <0 
\end{equation}
is strictly negative for $\forall  E \ne 0$. It has also been proven that the tracking error converges to zero as the time converges to infinity, ($\|E\| \longrightarrow 0$). The code below illustrates the symbolic proof in Isabelle theorem prover.   
\begin{mdframed}[style=MyFrame, frametitle ={}]	
	\begin{scriptsize}	
		\textcolor{royalblue(traditional)} {\textbf{theorem}} $Stb\_Eq\_36\_37:$ \\
		\textcolor{sapgreen} {\textbf{assumes}}  
		$"(set\_of\_definitions \ \ \phi \ \theta \ \psi \ \phi_d \ \theta_d \ \psi_d \ \eta \ \eta_d \ \eta' \ \eta_d' \ \eta'' \  \eta_d'' \ u \ \gamma \ e \ e' \ e'' \ \tau \ d \ d_{hat} \ E \ v \ \delta \\  C \ C_{hat} \ 
		N \ N_{hat} \ \Delta N \ \Delta d \ A \ B \ W \ Q \ P \ K_{\eta} \ K_r \ J \ J_{hat}"$ 
		\textcolor{sapgreen} {\textbf{and}} 
		$"assump1 \ d \ d_{hat} \ \Delta d"$
		\textcolor{sapgreen} {\textbf{and}} 
		$"assump3_a \ \eta_{d}''"$
		\textcolor{sapgreen} {\textbf{and}}
		$"assump3_b \ J \ J_{hat} \ \phi \ \theta \ \psi"$
		\textcolor{sapgreen} {\textbf{and}} 
		$"\forall E. \ E \ne 0"$
		\textcolor{sapgreen} {\textbf{and}}
		$"Lyapunov \ V \ E"$ 
		\textcolor{sapgreen} {\textbf{and}}
		$"V(E) = E \bullet (Q *_v E)"$
		\textcolor{sapgreen} {\textbf{and}} 
		$"(\forall t\in T. \ ((\lambda t. \ V(E)) \ \ has\_derivative \ 
		(\lambda t. \ V'(E))) (at \ t \ within \ T))"$	\ 
		\textcolor{sapgreen} {\textbf{and}}
		$"\eta'' = u - v + matrix\_inv(J) *_v  \gamma"$
		\textcolor{sapgreen} {\textbf{and}}
		$"E' = A *_v E + B *_v (v - (matrix\_inv(J) *_v  \gamma))"$	
		\textcolor{sapgreen} {\textbf{and}}		
		$"V'(E) = - (E \bullet (P *_v E)) + 2 * (((E v^{*} Q) v^{*} B) \bullet (v - matrix\_inv(J) *_v \gamma))"$ \\			
		\textcolor{sapgreen} {\textbf{shows}}
		$"norm(transpose(B) *_v (Q *_v E)) \ge \sigma \Longrightarrow V'(E) < 0"$ \\ 
		\textcolor{sapgreen} {\textbf{and}}	
		$"norm(transpose(B) *_v (Q *_v E)) < \sigma \Longrightarrow V'(E) < 0"$
		\textcolor{sapgreen} {\textbf{and}}	
		$"((\lambda t. norm(E)) \longrightarrow  0) \\  (at \ t \ within \ T)"$ \\
		\textcolor{royalblue(traditional)} {\textbf{proof -}} \\
		\textcolor{bleudefrance} {\textbf{show}} 
		$"norm(transpose(B) *_v (Q *_v E)) \ge \sigma \Longrightarrow V'(E) < 0"$ \
		\textcolor{royalblue(traditional)} {\textbf{using}}	
		$assms \ \ Eq\_22 \ \ rel\_simps(93)$ \
		\textcolor{royalblue(traditional)} {\textbf{by}} 
		$metis$	\\
		\textcolor{royalblue(traditional)} {\textbf{then}} \
		\textcolor{bleudefrance} {\textbf{show}}
		$"norm(transpose(B) *_v (Q *_v E)) < \sigma \Longrightarrow V'(E) < 0"$	\	
		\textcolor{royalblue(traditional)} {\textbf{using}} \		 
		$assms \ \ Eq\_22$\\ $ \ \ rel\_simps(93)$ \
		\textcolor{royalblue(traditional)} {\textbf{by}} 
		$metis$	\\
		\textcolor{bleudefrance} {\textbf{show}} \ 
		$"((\lambda t. norm(E)) \longrightarrow  0) (at \ t \ within \ T)"$ \
		\textcolor{royalblue(traditional)} {\textbf{using}} \		 
		$assms$ 
		\textcolor{royalblue(traditional)} {\textbf{by}}
		\textcolor{royalblue(traditional)} {\textbf{auto}} \\
		\textcolor{royalblue(traditional)} {\textbf{qed}}$ $
	\end{scriptsize}
\end{mdframed}  

\

\subsection {Onboard Verification for a Safe Flight using MetiTarski prover}
The control system of the UAV can be designed, simulated and verified at the model/design stage. The designed controller then formalised to its corresponding code and implemented into the autopilot system, which controls the aircraft trajectory. The aircraft controlled by the autopilot can be exposed to gusts of wind which may cause unstable flight. In this case, the autopilot system cannot be informed if the aircraft has entered an unstable region which may cause a crash or lose of human(s) life. Therefore, we proposed using of ATP tool represented by MetiTarski prover for onboard verification that the prover can check the stability state of the aircraft and inform the autopilot in case of any unstable behaviour detected. 

The autopilot then can send warning messages to the user/pilot or base station to perform, for instance, an emergency safe landing using autolanding techniques such as in \cite{c88e}. This will ensure more safe flight and may avoid losing the aircraft or any harm to humans and properties. We have chosen MetiTarski ATP to verify the controller stability of the aircraft due to its ability of deal with inequalities with numerical real numbers. Unlike the previous verification stage using Isabelle, MetiTarski proves the statements automatically without the need to any interaction with designer/engineer. 

\begin{figure}[thpb]
	\centering
	\includegraphics[scale=0.5]{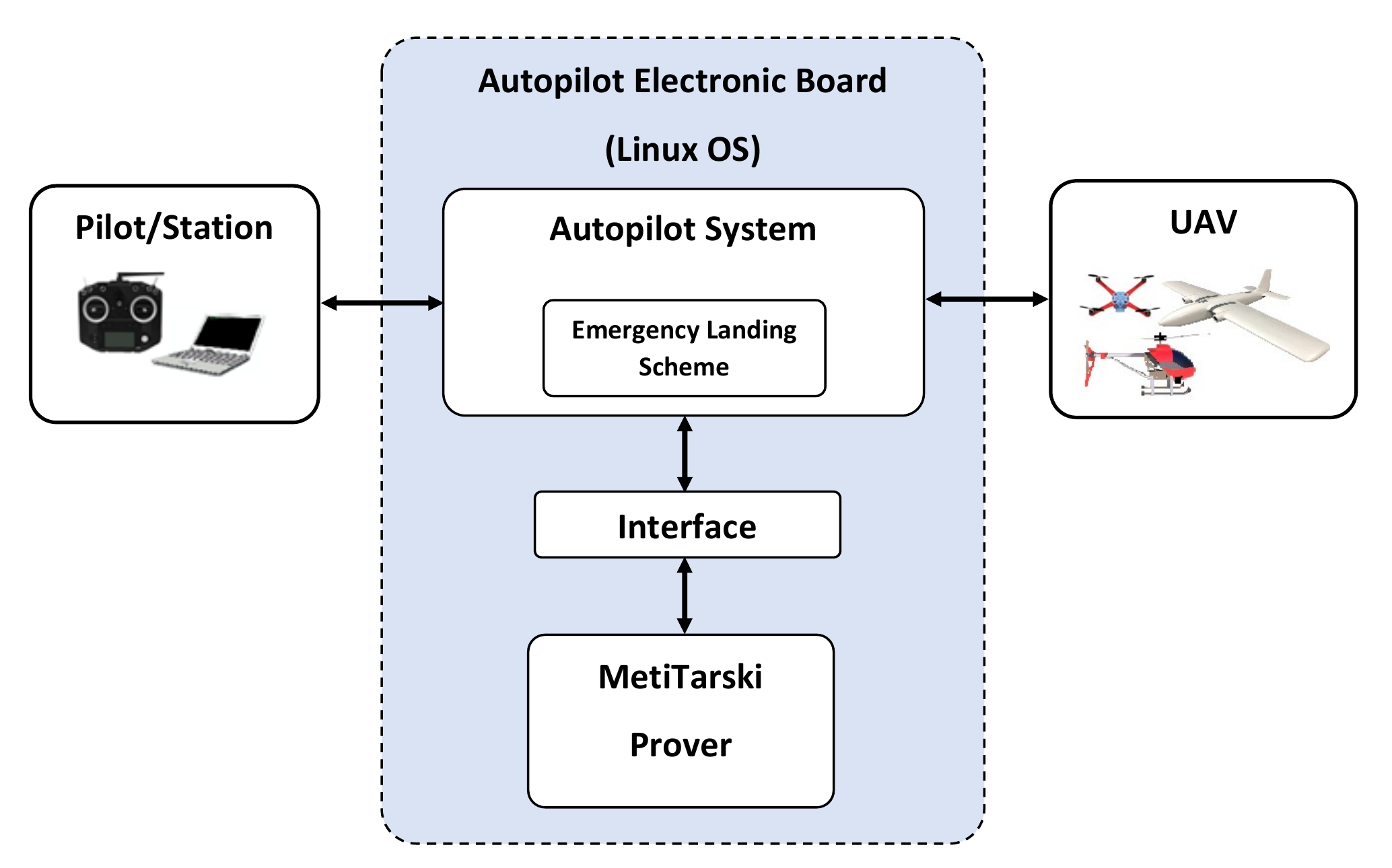}
	\caption{Onboard verification framework of UAVs}
	\label{fig2}
\end{figure}

MetiTarski can be implemented on the autopilot's electronic board such as PixHawk \cite{c88g} or Navio2 \cite{c88h} Raspberry Pi. These electronic autopilots use the Linux operating system where  MetiTarski can also be installed. Therefore, an interface between the two systems (autopilot and MetiTarski) is easy to create in practice. The proposed onboard framework is illustrated in Fig.\ref{fig2}. The applicability of this approach is illustrated  in simulation by interfacing the Simulink/Matlab model with MetiTarski and test system stability. Note that the quadcopter that has been implemented in Simulink considers the nonlinear dynamics of the craft based on the MathWorks model in \cite{c86}, which is widely used.

Considering the stability analysis stated in Eqs. \eqref{FV_eq39} and \eqref{FV_eq39a}, the time derivative of the Lyapunov function will be tested to check whether it is negative definite or not. If it is not negative definite, then this indicates that the control system is out of its stability region, hence the autopilot can pass warning messages to the pilot or station to take an action or perform an emergency landing. The verification process starts by formalising the stability equations \eqref{FV_eq39} and \eqref{FV_eq39a} into a FOL syntax. The parameters of these equations are passed from the autopilot system to MetiTarski prover via an interface scheme. Afterwards, the test is conducted in the MetiTarski prover such that $\dot{V}(E) < 0$. We have simulated the above procedures in Simulink/Matlab to illustrate its applicability. The stability equations, Eqs. \eqref{FV_eq39} and \eqref{FV_eq39a}, are simplified using symbolic computations in Matlab before formalising them into FOL syntax. The parameters included in both stability equations are passed from Simulink/Matlab to MetiTarski to perform the monitoring test.  The following code shows an example of stability check in MetiTarski prover for Eq. \eqref{FV_eq39} (note that the full code can be found in our online  repository):
\begin{mdframed}[style=MyFrame, frametitle 	={}]	
	\begin{scriptsize}	
		$fof(Stability\_Eq15,conjecture, ![E\_1,E\_2,E\_3,E\_4,E\_5,E\_6,Phi,Theta] : ?[V\_1,V\_2,V\_3,\\ Delta\_E]: \\ 
		\% assumptions \\
		( E\_1  = 1.6 \ \& \ E\_2  = 3.1 \ \& \ E\_3 = 2 \ \& \ E\_4  = 9.3 \ \& \ E\_5 = 6.8 \ \& \ E\_6  = 4.8 \\
		\& \ Phi > -1.5708 \ \& \ Phi <1.5708 \ \& \ Theta > -1.5708 \ \& \ Theta <1.5708 \\
		\& \ abs(V\_1) <= (0.5*(1.2+(0.004*abs(E\_4))+(17.5*abs(E\_1))) + (173*(0.001+0.001))) \\
		\& \ abs(V\_2) <= (0.5*(1.2+(0.004*abs(E\_5))+(17.5*abs(E\_2))) + (173*(0.001+0.001))) \\
		\& \ abs(V\_3) <= (0.5*(1.2+(0.482675*abs(E\_6))+(1.8*abs(E\_3))) + (173*(0.001+0.001))) \\
		\& \ Delta\_E > 0 \ \& \ Delta\_E >= sqrt(V\_1^2+V\_2^2+V\_3^2)/170.5 \\
		\% implies \\
		=> .... <0 )).$			 
	\end{scriptsize}
\end{mdframed}  

\noindent and for Eq. \eqref{FV_eq39a},
\begin{mdframed}[style=MyFrame, frametitle 	={}]	
	\begin{scriptsize}	
		$fof(Stability\_Eq16,conjecture, ![E\_1,E\_2,E\_3,E\_4,E\_5,E\_6,Phi,Theta] : ?[V\_1,V\_2,V\_3, \\ Delta\_E]: \\
		\% assumptions \\
		(E\_1  = 2.9 \ \& \ E\_2  = 1.2 \ \& \ E\_3 = 1.8 \ \& \ E\_4  = 6.9 \ \& \ E\_5 = 10.5 \ \& \ E\_6  = 5 \\
		\& \ Phi > -1.5708 \ \& \ Phi <1.5708 \ \& \  Theta > -1.5708 \ \& \ Theta <1.5708 \\
		\& \ abs(V\_1) <= (0.5*(1.2+(0.004*abs(E\_4))+(17.5*abs(E\_1))) + (173*(0.001+0.001))) \\
		\& \ abs(V\_2) <= (0.5*(1.2+(0.004*abs(E\_5))+(17.5*abs(E\_2))) + (173*(0.001+0.001))) \\
		\& \ abs(V\_3) <= (0.5*(1.2+(0.482675*abs(E\_6))+(1.8*abs(E\_3))) + (173*(0.001+0.001))) \\
		\& \ Delta\_E > 0 \ \& \ Delta\_E >= sqrt(V\_1^2+V\_2^2+V\_3^2)/170.5 \\
		\% implies \\
		=> .... <0 )).$			 
	\end{scriptsize}
\end{mdframed}

\section{Conclusions}
This paper has introduced a new verification framework for safety-critical control systems  by applying  the power of  higher-order-logic-based interactive theorem provers and a first-order logic-based automated theorem prover to verify the control system of unmanned aerial vehicles and to ensure aircraft control system stability and performance. The framework relies on two stages, the first is for verifying the design of the control system and its stability and the second is for onboard monitoring of the aircraft's stability to ensure flight safety. 
The framework has been demonstrated on a robust attitude controller of a generic quadcopter UAV to verify the correctness of the design and stability analysis in addition to onboard monitoring the conditions of its dynamical stability while the aircraft is flying. The aircraft's attitudes are controlled by a nonlinear robust controller, which is designed using  inverse dynamics control and it takes into account dynamical uncertainty and external disturbances.

The methods used in the verification stages go significantly beyond symbolic computation of inequalities for the Lyapunov theory as concepts
of convergence as mappings of functions and quantifications over sets of functions are used in Isabelle and as such they were not be found in prior literature in aviation control systems.

The paper illustrated the applicability and the power of interactive theorem provers relying on HOL  and automated theorem prover represented by FOL for control theory.   This is promising and may encourage the use of such methods in  control system verification 
of safety-critical systems in general.   The symbolic methods are generic and potentially generalise to verification of a variety of
industrial control systems, where performance loss is damaging and therefore  analysis is important to be carried out formally.



  \bibliographystyle{elsarticle-num} 
  \bibliography{bibliography}





\end{document}